\documentclass[12pt]{article}
\usepackage[right=1.25in,left=1.25in,top=1.1in,bottom=1.1in]{geometry}
\usepackage{hyperref}
\hypersetup{colorlinks, citecolor=blue, filecolor=blue, linkcolor=blue, urlcolor=blue}
\usepackage{graphicx}
\usepackage{url}
\usepackage[round]{natbib}
\usepackage{amsmath,amsthm} 
\usepackage{engord}
\usepackage{float}
\usepackage{subfig}
\usepackage{pdflscape}
\usepackage{booktabs}
\usepackage{pgfplots}
\pgfplotsset{compat=1.14}
\pgfplotsset{every axis label/.append style={font=\tiny}}
\usepackage[labelsep=period]{caption} 

\usepackage{amssymb} 
\usepackage{multirow} 

\usepackage{xr}

\usepackage{setspace}
\usepackage{sectsty}
\sectionfont{\large}
\subsectionfont{\normalsize}
\subsubsectionfont{\normalsize}

\newcommand{\specialcell}[2][c]{\begin{tabular}[#1]{@{}l@{}}#2\end{tabular}}

\title{ \vspace*{-2.5cm} \hspace*{-0.5cm}Discrimination and Constraints: Evidence from The Voice}

\author{Anuar Assamidanov\thanks{ Department of Economics,  Claremont Graduate University, 150 E 10th St, Claremont, CA 91711. \href{mailto:anuar.assamidanov@cgu.edu}{anuar.assamidanov@cgu.edu} \\ I am grateful for valuable comments and discussions from Gregory DeAngelo, Fernando Lozano, Mark Hoekstra, CarlyWill Sloan, Bryan McCannon, Benjamin Blemings, Aruna Ranganathan and numerous seminar participants. Any errors are my own.
} }

\date{ \vspace*{0.5cm} November 2022\\
} 

\begin{document}

\bgroup
\let\footnoterule\relax

\begin{singlespace}
\maketitle

\begin{abstract}
    \noindent 
    
    Gender discrimination in the hiring process is one significant factor contributing to labor market disparities. However, there is little evidence on the extent to which gender bias by hiring managers is responsible for these disparities. In this paper, I exploit a unique dataset of blind auditions of \emph{The Voice} television show as an experiment to identify own gender bias in the selection process. The first televised stage audition, in which four noteworthy recording artists are coaches, listens to the contestants “blindly”  (chairs facing away from the stage) to avoid seeing the contestant. Using a difference-in-differences estimation strategy, a coach (hiring person) is demonstrably exogenous with respect to the artist's gender, I find that artists are 4.5 percentage points (11 percent) more likely to be selected when they are the recipients of an opposite-gender coach. I also utilize the machine-learning approach in Athey et al. (2018) to include heterogeneity from team gender composition, order of performance, and failure rates of the coaches. The findings offer a new perspective to enrich past research on gender discrimination, shedding light on the instances of gender bias variation by the gender of the decision maker and team gender composition.

\end{abstract}
\end{singlespace}
\thispagestyle{empty}

\clearpage
\egroup
\setcounter{page}{1}



Gender discrimination in the hiring process is one significant factor contributing to the poor labor market \citep{Blau2007}. Gender differences in hiring are challenging to measure, making it difficult to distinguish its effect on employment from confounding variables, such as differences in human capital and relevant skills. However, \citet{Baert2018} have cataloged studies on evidence of gender discrimination in hiring using experimental methods where confounding factors were disentangled to estimate the true effect on the labor market outcomes.

Studies have found  hiring discrimination practices against women in the labor market, but most of this research has utilized experimental methods\citep{Bertrand2017, Baert2018, Neumark2018} .  Within these experiments, pairs of fictitious job applications, only differing by the gender of the candidate, are sent to actual job openings. Discrimination is identified with the subsequent call-back from the employer and the gender of the candidate. The correspondence testing methodology is the gold standard for estimating hiring discrimination in the labor market \citep{Baert2015}. However, in this hiring literature, the gender of hiring decision-makers is unobservable, or decisions are made collectively. Furthermore, these experiments only capture call-back rates and do not go beyond the initial stage of the hiring process. In addition, assuming employers’ hiring decisions are based on a cut-off rule, there could be unobserved variable differences in the productivity between the two groups, which can cause biased discrimination measures \citep{Neumark2012}. 

Another part of the literature on gender discrimination is own gender bias (i.e., favoritism toward people of one's own gender) in hiring decisions. Laboratory experiments often find that women show own gender bias when information is processed automatically, but these results are not found for men \citep{Rudman2004}. However, it is unclear whether these results hold in real-world hiring decisions, which are likely to be characterized by a reflective (non-automatic) process. The existing studies on real-world hiring decisions typically consider particular segments of the labor market and find mixed evidence for own gender bias \citep{Booth2010, BAGUES2010, Bagues2017}. Because of the data limitations in these settings, it is hard to distinguish if there is favoritism of the gender of the employer over the gender of the candidate (i.e., own-gender bias). This paper's primary purpose is to test whether own-gender bias exists in the hiring process. Potential findings could inform our understanding of the current labor market at the individual level and the subsequent large-scale implications. 
 
The main difficulty in testing gender bias in the hiring process is that hiring decision-makers and applicants are not the outcome of a random selection process to claim a causal factor in generating own gender bias. Also, the fundamental identification challenge is that researchers only observe real-world data on already hired workers—if the applicant is not employed, it is never recorded. Due to these limitations, there is an open question in this literature if the gender of the hiring decision-maker determines the gender of applicants being hired. In this paper, I exploit a unique dataset of blind auditions of the Voice TV show as a natural experiment to identify the own-gender bias in the selection process.

I assess own-gender bias in The Voice show, a setting in which a coach (hiring person) is demonstrably exogenous with respect to the artist's gender. The first televised stage in the Voice TV show is the blind auditions. The four coaches, all noteworthy recording artists, listen to the contestants in chairs facing away from the stage to avoid seeing the contestants. If a coach likes the contestant's voice, they press a button to rotate their chairs to signal that they are interested in working with that contestant. The advantage of these estimations is that the coaches only observe the voice of the contestants and conclude from the voice the probable gender of the contestant. So, due to the "blind" decision process, other characteristics related to the artist and coach will be eliminated in this setting. I believe this is a plausible environment to use a generalized difference-in-difference identification strategy with random assignment of coaches to participants.  Like \citet{Price2010} study on racial bias by NBA referees, I compare differences in the probability of female and male contestants being chosen by female and male coaches.
 
By drawing several types of inferences, the analysis shows systematic evidence of an opposite-gender bias (favoritism toward the opposite gender). Coaches that are men predominantly prefer contestants that are women, while female coaches prefer male contestants. Contestants are 4.5 percentage points (11 percent) more likely to be selected by a coach that is the opposite gender. To examine heterogeneity more systematically, I adapt the machine-learning approach of \citet{Athey2019}. I have extended the analysis by including heterogeneity in the gender composition of each coach's team, the coaches' failure rates, and the performance order. The results provide intriguing insights into gender bias that varies widely across team composition during the selection of new contestants. 

This paper offers one of the first explorations of hiring practices using random variation in gender to examine the effects on the selection process. The random variation in gender enables me to overcome potential concerns that observed disparities could be due to unobserved differences across contestant gender. Second, this paper applies the \citet{Athey2019} method of causal forest machine learning inference on heterogeneous treatment effects to a quasi-experimental difference-in-difference identification strategy. 

This paper is closely related to work by \citet{Carlsson2019}. Their work provides related evidence suggesting a role for in-group gender preferences has been documented in many other contexts. They investigated in-group and own-gender bias in real-world hiring decisions by combining data on the gender of the recruiter and the share of women employees in many firms with data from a large-scale field experiment on hiring. The results suggested that women (female recruiters or firms with a high share of female employees) favor women in the recruitment process. Nevertheless, this study provides only a tiny fraction of the participating firms' information regarding recruiters' gender, which might result in a considerable measurement error. 

Similar to this work, \citet{Goldin2000} studied blindness in the orchestra context to examine sex-biased orchestra member selection as a policy intervention. This paper differs from their work in that I can observe the gender of the hiring and contestant, and coaches can observe the gender of the performer based on their voice. The coaches' decisions will also be made individually, whereas, in the blind audition in the orchestra, there was a collective decision. Finally, the other uniqueness of this setting is that I could analyze the market structure where each coach could be assumed as a firm. I was able to illustrate the market power of the coaches by leveraging the order of the performance and the genre of the song they performed. Controlling for all these factors gives a new dimension to gender discrimination in the competitive market.

The rest of the paper is organized as follows. Section \ref{sec:background} provides a brief background on The Voice, summarizes the blind audition stage of the show, and gives a preview of our estimation strategy. Section \ref{sec:data} describes the data. Section \ref{sec:results} describes the empirical strategy in detail. Section \ref{sec:conclusion} presents results, and Section \ref{sec:conclusion} concludes.

\section{Background 
\label{sec:background}}

An international television singing competition franchise initially created by Dutch television producer John de Mol and singer Roel van Velzen. One hundred forty-five countries have adopted the format and begun airing their versions since 2010. The show's format features five stages of competition: producers' auditions, blind auditions, battle rounds, knockouts (since 2012), and live performance shows. 

Each season employs a panel of four coaches who critique the artists' performances and guide their teams of selected artists through the remainder of the season. They also compete to ensure that their act wins the competition, thus making them the winning coach. While these constraints mean that the assignment of coaches to the season is not literally random, the more relevant claim for this research design is an entirely arbitrary assignment of coaches to the artists.

Each season begins with the "Blind Auditions," where coaches form their team of artists they mentor through the remainder of the season. The coaches' chairs are faced toward the audience during artists' performances; those interested in an artist press their button, which turns their chair towards the artist and illuminates the bottom of the chair to read "I want you." Each coach has the length of the auditioner's performance (about one minute) to decide if he or she wants that singer on his or her team; if two or more coaches want the same singer (as happens frequently), the singer has the final choice of coach. After the performance, an artist either defaults to the only coach who turned around or selects the coach if more than one coach expresses interest. The coaches not seeing the artists is a setting in which coaches' encounters are demonstrably exogenous with respect to artists' performances \citep{Voice}.

\section{Data \label{sec:data}}
I used the blind audition part of the show as a natural experiment in this work. The data on a blind audition was compiled from historical Voice TV shows on Wikipedia. I web-scraped the results of the blind audition for four countries: the UK, France, Germany, and Australia. Each country held approximately ten seasons each year, starting from the year 2012. Each country's Voice Wiki page is divided by season, each season is divided by the episodes, and the order of the performance splits each episode. The page includes the singer and coach's information, the singer's age, the order of performance, and the song. 

From the Wiki page, I constructed the outcome variable based on the coaches who pressed the button if the coach chose the specific artist. Thus, I can build the result of the four coaches for each artist. Similarly, the page provides the actual order of the performance. Since the primary goal of the analysis is to identify gender-specific preferences, I needed to determine the gender of the artist. 

I infer the gender of the artists based on their first names. I used an online application programming interface (API) called Genderize.io. The application has pre-trained machine learning algorithms that are used to predict gender. The API gives a confidence level in probabilistic terms and the exact name count in their already used datasets. Utilizing a threshold of 90{\%} probability, I am able to attach a gender to approximately 97{\%} of the whole dataset. For the remaining 3{\%} of the first names, I hardcoded by checking the performance on the YouTube webpage. The gender of the coaches can easily be identified as all of them are well-established artists.
 
I also used the genre of the song the artist performs. Using song information from the Wikipedia page, I extracted the genre of the music from the Spotify API. The API gives multiple genres that correspond to the given song. By doing frequency distribution, I determined the most frequent genres and came up with the eleven unique genres. 

Summary statistics are shown at the coach level in Table \ref{tab:summary_statistics}. Column (1) includes all coaches in the sample, while Columns (2) and (3) disaggregate to coaches who are female or male, respectively. The data has 11,972 observations, 58 unique coaches, and 3,009 unique artists. 31.6 percent of the coaches were female, though there is considerable heterogeneity across the team gender composition and failure rates for female and male artists. Similarly, about 46.3 percent of artists are men, providing decent statistical power for an own-gender bias analysis.  

My primary analysis includes all the coaches' choices, with approximately 40 percent of the artists being chosen in the blind audition round. Female artists are chosen about 21.6 percent of the time. The difference between female coaches and male coaches choosing female artists is a 1 percent difference. This pattern of having a 1{\%} difference continues to hold for the male artists who are chosen 18 percent of the time. Thus, a naive comparison of the raw data for female and male gender would yield a conclusion of no gender bias. Since the preference of female and male artists across female and male coaches is similar, the selection difference lies at about one percent. To justify this premise, I examine the order of each performance at the artist level and how TV show-related specification affects decision-making.

Table  \ref{tab:summary_statistics} also provides summary statistics for TV show-related specifications, considering them heterogeneous in the analysis. Each coach and artist has a unique specification, such as the order of their performance, their preference towards a specific coach, and the genre of the song.  Ultimately, 16.8 percent of the artists ended up on a team. From that team configuration, we can determine the gender composition of the team, which is the number of women and men on the team during the blind audition stage. The average difference between the number of males and females for each order on the team is -0.533, meaning that the number of female artists is 0.5 people higher on average compared with the male artists. Additional information is obtained when two or more coaches pursue the same artist, providing the artist with the final choice. I estimated the failure rate for these cases, and each coach has about a 0.27 failure rate for female artists and 0.24 for male artists. This is the rate at which a coach wanted an artist of a particular gender on their team, and the artist did not choose the coach. Both of these variables have considerable differences across the gender of the coach, which enriches our study to identify own-gender bias in the selection process.

\begin{table}[H]
    \begin{center}
    \caption{Summary statistics for coaches}
    \begin{tabular}{lcccc} 
    \toprule
    & All coaches & Male coaches & Female coaches \\
    & (N = 58) & (N = 36) & (N = 22) \\
    & (1) & (2) & (3) \\
    \midrule
   \\
    Female Coaches & 0.684 & 1 &	0 \\
    \\
    Male Artist
 Auditioned & 0.463	& 0.465	& 0.458 \\
    \\
    Coaches’ Choice in the Audition & 0.398 &	0.399 &	0.394  \\
    \\
    Female Artist Choice of Coach & 0.216 &	0.219 &	0.209 \\
    \\
    Artist’s choice of coach & 0.168 &	0.168 &	0.169 \\
    \\
    Male relative to female & -0.533 & -0.514 &	-0.574  \\
    in the team & (2.57) & (2.55) & (2.60) \\
    \\
    The failure rate & 0.273 &	0.238 &	0.289\\
    for Female Artists & (0.148) &	(0.151) &	(0.134) \\
    \\
    The failure rate & 0.24	& 0.234	 & 0.242\\
    for Male Artists & (0.141)	& (0.145)	 & (0.134) \\
    \\
    Number of Observations & 11,972	& 8,189	& 3,783 \\
    \bottomrule
    \end{tabular}
    \label{tab:summary_statistics}
    \end{center}
    
	\begin{minipage}{0.90\textwidth} 
	{\footnotesize 
    Notes: This table presents summary statistics for the estimation sample. The unit of observation is performance in the blind audition stage. The table shows statistics of all performances for which artists and drivers are men and women. The sample includes four countries, and each country has ten seasons. Standard deviations in parentheses. All values reported in this table are unconditional average rates of the variables indicated in rows.
	\par}
	\end{minipage}
\end{table}

As discussed and shown in the Section \ref{sec:background}, the institutional setting provides for the exogenous assignment of gender encounters between coach and artist. Table \ref{tab:summary_statistics_2} examines this exogeneity assumption by estimating differences in the observed characteristics of male and female artists. The first column shows mean characteristics for all artists, while the second and third columns present average outcomes for male and female artists. The fourth column reports the difference between the two means.  Comparing overall raw means is consistent with the assignment procedure described above. There is no significant difference between male and female artists in any show-related specifications, such as being chosen by coaches, selecting a team, being chosen by a male coach, and the order of their performance. This supports our identification strategy: in a given show setting, there is little evidence to suggest that male artists' performances systematically differ from those performed by female ones. However, I also find some statistically significant differences. For example, male artists are 5.2 percent more likely to perform pop genre songs. Similarly, the share of classic style and the share of rock style is lower, while the share of country style is higher among male artists. All these significant observables will be estimated in the setting as confounding factors. 

\begin{table}[H]
    \begin{center}
    \caption{Balancing tests for the performance of artists}
    \begin{tabular}{lccccc} 
    \toprule
    & All Artists & Male Artist & Female Artist & Difference in Means\\
& & & & Male vs. Female Artist\\
    & (1) & (2) & (3) & (4)\\
    \midrule
   \\
    Coaches’ Choice  & 0.398 &	0.402 &	0.393 &	0.009 \\
    in the Audition& & & & [.009] \\
    \\
    Artist’s choice of coach & 0.168 &	0.17 &	0.167 &	0.003 \\
     & & & & [0.007] \\
 \\
    Auditioned by Male Coach & 0.684 & 0.681 &	0.687 &	-.005 \\
& & & &[0.009] \\
    \\
    Performance Order & 42.19 &	42.34 &	42.02 &	0.328 \\
&(25.49) & (25.61) & (25.39) &[0.468]\\
    \\
    Pop & 0.557 &	0.581 &	0.529 &	0.052*** \\
& & & & [0.009] \\
    \\
    Hip-hop & 0.121	& 0.122 & 0.119	& 0.004 \\
& & & & [0.006] \\
    \\
    Rock & 0.27	& 0.233 &	0.314 &	-0.082*** \\
& & & & [0.008] \\
    \\
    Country & 0.046	& 0.050	& 0.042	& 0.009** \\
& & & & [0.004]\\
 \\
    \\
    R{\&}b  & 0.0695 &	0.075 &	.062 &	0.013 \\
& & & & [0.005] \\
    \\
    Funk  & 0.0736 & 0.07 & 0.0776	& -0.007 \\
& & & & [0.005]\\
    \\
    Jazz  & 0.0312 &	0.0331 &	0.029 &	0.004 \\
& & & & [0.003] \\
    \\
    Number of Observations & 11,972 & 6,432	& 5,540 & 11,972\\
    \bottomrule
    \end{tabular}
    \label{tab:summary_statistics_2}
    \end{center}
    
	\begin{minipage}{0.90\textwidth} 
	{\footnotesize 
    Notes: Standard deviations in parentheses in columns 1,2 and 3. Each entry in column 4 is derived from a separate OLS regression where explanatory variable is indicator for male artist.  * \(p<0.1\), ** \(p<0.05\), *** \(p<0.01\)
	\par}
	\end{minipage}
\end{table}

\section{Methodology \label{sec:methodology}}

The identification strategy used in this study is Difference in Differences with Dummy Outcomes \citep{Price2010}, which focuses on coach and artist gender interaction. Specifically, we estimate the following linear probability model: 

\begin{equation}
 \begin{array}{c}
    Y_{it} = \alpha_i + \beta_1 MaleCoach_c + \beta_2MaleArtist_i + \beta_3 (MaleArtist\ {\&}\ MaleCoach)_{i,c} \\ + \ Coach_c+ X_{i,c} + \epsilon_{it}
\end{array}
    \label{eq:ols}
\end{equation}

where the outcome of interest, $Choice_{i,c}$   is the binary variable equal to one when the coach chooses the artist in performance, where $i$ denotes an artist selected by the coach $c$ . Similarly, $MaleCoach_{c}$ is one when the Coach is male and zero otherwise. $MaleArtist_{c}$ is one when the artist is male and 0 otherwise. The primary variable of interest  is the interaction between male coaches and male contestants.  $\beta_2$  captures the difference between male artists chosen by female coaches relative to female artists chosen by female coaches,  allowing for artists to differ in their overall propensities to be selected. $\beta_1$ coefficient captures differences in selection rates of female artists by male coaches relative to female artists by female coaches, allowing coaches to differ in their overall propensities to select the artist.  $\beta_3$ is the difference in differences coefficient of interest and indicates the degree of own-gender bias. Notably, the difference in differences estimator provides only a symmetric and relative measure of gender bias.  So the coefficient captures the comparison of the difference in selection rates of female artists by male coaches relative to female artists by female coaches to the difference in selection rates of male artists by male coaches relative to male artists by female coaches.   If there is no gender bias, then  $\beta_3$ = 0. A positive sign indicates same-gender leniency (or other-gender harshness) in blind audition selection.

The richness of the data allows for extending this analysis to control for the various confounding factors such as  contestant-specific characteristics and coach fixed effect, genre, and order of the performance. In the model, I include $Coach_{c}$ coach-specific fixed effect, which captures time-invariant coach characteristics. $X_{i,c}$ is the set of control variables at the performance level, including the genre of song, performance order, the number of male relative to female artists in the coach's team, the coach's failure rate for male and female artists during the audition as well as fixed effect for the country of the TV show. I also cluster standard errors for all estimations at the coach level, as this seems ex-ante to be the appropriate cluster for most autocorrelation concerns.

The regression model in the difference-in-differences framework does not capture potential heterogeneous effects across covariates. There could be a positive effect on some groups in the covariates but a negative or no effect on other groups. For example, the coaches in the initial stage of the blind audition could elicit own-gender bias, but in the later stage, opposite-gender bias. The standard regression model will provide a weighted average of effect size, even if an effect varies across a population represented by different groups. One plausible solution is to estimate heterogeneous treatment effects. Instead of just estimating one effect, I estimate a distribution of effect size for a given unit with a given set of attributes and what their impact might be. To analyze this heterogeneous effect of own-gender bias, I incorporated the first difference approach into the causal forest approach. I will discuss the performance of the proposed causal forest approach and how to apply it to my empirical example. 

The causal forest approach is consistent and asymptotically Gaussian and provides an estimator for their asymptotic variance that allows us to construct valid confidence intervals \citep{Athey2019}. Causal forest utilizes a forest-based approach to calculate a similarity weight and then applies the local generalized method of moments to estimate  based on a weighted set of neighbors. The authors also show that using the centered instead of the original outcome and treatment allows a forest to focus its splits on the features that affect the treatment effect instead of the primary outcome. Intuitively, a causal forest consists of regression trees \citep{Athey2016, breiman2017classification}, which estimate conditional average treatment effects (CATE) at the leaves of the trees instead of predicting the outcome variables.

I utilize a two-step approach by incorporating the first-difference approach into the causal forest approach, the similar two-step approach in the standard difference-in-difference method. In the first step, I estimate the first difference, which is the change in the selection of individual contestants from a male coach to a female coach. This step teases out the systematic differences between coaches in female and male coach groups. In the second step, I use the outcome variables obtained from the first step and apply the causal forest approach for heterogeneous treatment-effect analysis. This step teases out the effect of contestant gender trends that exist in both women and men.

My methods rely heavily on the causal forest approach, a machine learning method that can estimate heterogeneous causal effect functions under the above assumptions. The main strength associated with the causal forest is that it provides a strategy for learning patterns of heterogeneity from the data. This method requires little researcher discretion compared to traditional subgroup analyses. For instance, one does not have to decide which cutoffs to use to analyze heterogeneity in effects along continuous covariates, nor do we require to define the functional condition of the heterogeneity; the causal forest algorithm is designed to automate this search and find the most appropriate model that best characterizes the heterogeneous effect in the data \citep{Bonander2021}.  

\section{Results\label{sec:results}}

\begin{figure}[H]
    \begin{center}
    \caption{Baseline Own-Gender Bias}
    \label{fig:mainsummary}
    \includegraphics[width=5.0in]{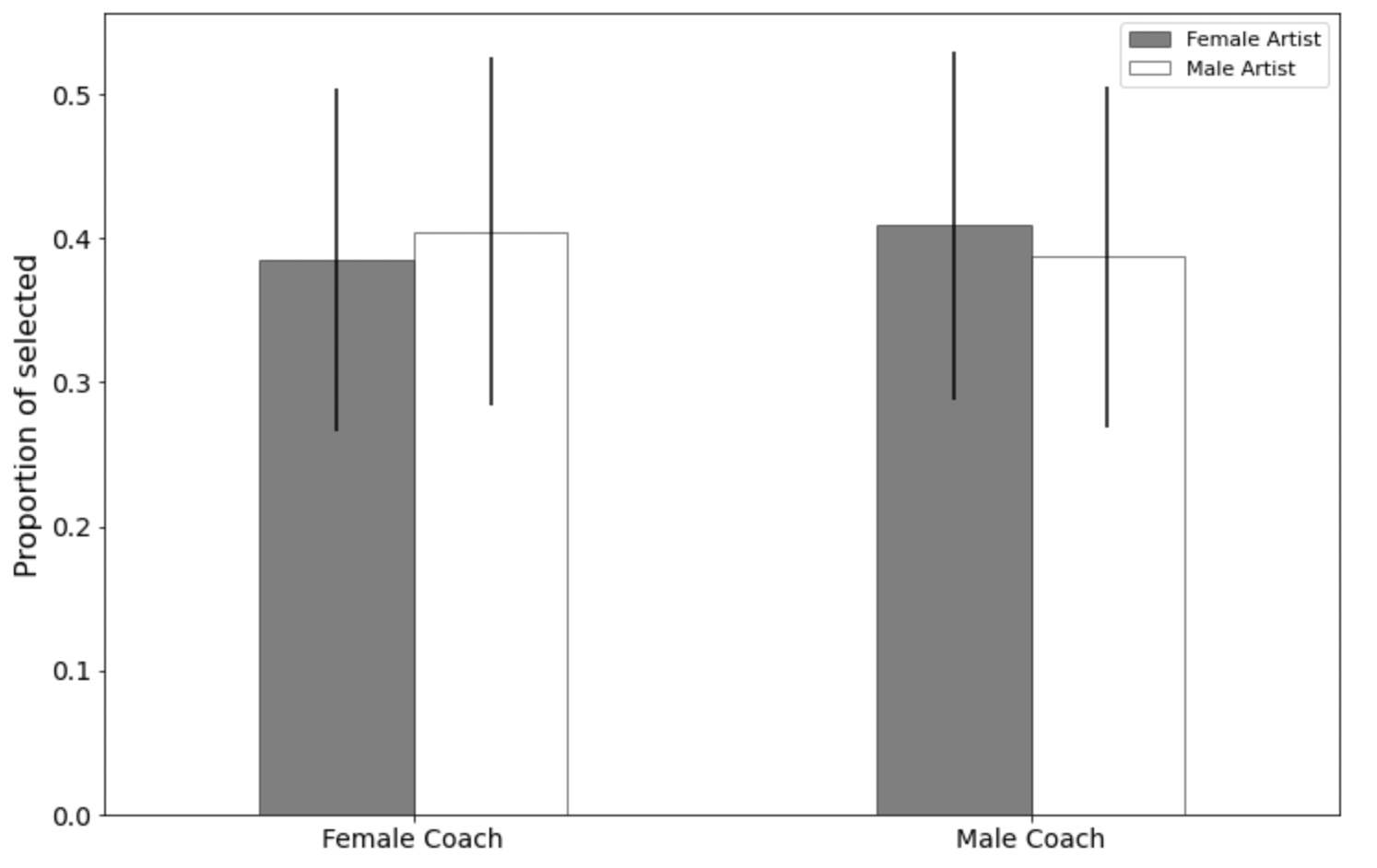}
    \end{center}
    \vspace{0.2cm}
    \begin{minipage}{0.95\textwidth} 
	{\footnotesize Notes: This figure reports the share of selected male and female artists assigned to male and female coaches. There is a statistical difference in the selection of female artists assigned to female or male coaches. The same is true for male artists.
	\par
	}
	\end{minipage}
\end{figure}

My identification strategy is to leverage the exogenous assignment of coaches to the artist and use a difference-in-differences estimator across these coach-artist interactions to estimate the degree of gender bias in selection. The baseline difference in differences results shows that female coaches are more likely to choose male artists, and male coaches are more likely to select female artists, opposite gender bias (Figure \ref{fig:mainsummary}). A simple calculation suggests an opposite gender bias of 4.2 percentage points. 

\subsection{Effect of Own-Gender Coaches on Selection \label{sec:heterogeniety}}

Table \ref{tab:regressions} reports results from estimating Equation 1 using four separate specifications. The "same gender coach" difference in differences coefficient shows how much more or less likely an artist is to be chosen by a coach of a different gender relative to one of the same genders as the artist, adjusted for overall gender propensities of artists to get selected and coaches to choose. According to Table \ref{tab:regressions}, the results represent the regression outcome of the chosen dependent variable artist. Overall, it can be seen that the primary independent variable interaction between the gender of the coach and artist is showing consistent output with controlling various confounders. The outcome gives ranges of - 4.4 and - 5 percentage points, indicating that magnitude increases with covariates. Standard errors also decreased and showed a range of 15{\%} and 9{\%} for the margin of error or alpha level.

Column (1) of Table \ref{tab:regressions}, includes coach and country effects and no other controls. The estimated degree of gender bias for selection in the male/female comparison is -4.4 percentage points. That is, after adjusting for gender-specific differences in propensities for contestants to perform and adjusting for gender-specific differences in coaches' propensities to select, a male contestant who is exogenously assigned a male coach is 4.4 percentage points less likely to be chosen relative to a male contestant who is exogenously assigned a female coach (and vice versa). Given a mean of 39.75 percent, the pooled result indicates that gender bias causes a contestant assigned an other-gender coach to be nearly 11 percent more likely to be selected for any performance.

\begin{table}[H]\centering
\def\sym#1{\ifmmode^{#1}\else\(^{#1}\)\fi}
\caption{Estimates of own-gender Bias for Artist Selection}
\begin{tabular}{l*{5}{c}}
\toprule
&\multicolumn{1}{c}{(1)}&\multicolumn{1}{c}{(2)}&\multicolumn{1}{c}{(3)}&\multicolumn{1}{c}{(4)}\\
\midrule
\\
\multicolumn{4}{l}{\textbf{Outcome: Choice}} \\
Male Coach*Male Artist &     -0.0441\sym{*} & -0.0453\sym{**} & -0.0485\sym{**} & -0.0498\sym{**}

  \\
  & (0.0237) & (0.0221) & (0.0216) & (0.0216)
    \\
  \\
Male Coach & 0.0226              & -0.212\sym{***}     & -0.277\sym{***} & -0.274\sym{***} \\
 & (0.0185) & (0.0170) & (0.0166) & (0.0166)
 \\
 \\
 Male Artist & 0.0144 & 0.0199 & 0.0206 & 0.0190\\

& (0.0185) & (0.0170) & (0.0166) & (0.0166)\\
\\
Mean       &     0.398         &     0.398          &   0.398  & 0.398    \\
$N$          & 11,195 & 11,195 & 11,195 & 11,135
  \\
  \\
  Coach FE       &     Y         &     Y          &   Y  &  Y    \\
  Country FE       &     Y         &     Y          &   Y  &  Y    \\
Performance Order FE          & - & Y & Y & Y
  \\
  Genre FE          & - & - & Y & Y\\
  Constraints  & - & - & - & Y
  \\
\bottomrule
\end{tabular}
\label{tab:regressions}
\vspace*{0.09cm}
\begin{minipage}{0.95\textwidth} 
{\footnotesize 
Notes: Table 3 presents a difference in differences estimate of gender bias using regression of Equation 1. This table reports the coefficient on the interaction of Male Artist and Male Coach from the regression of Choice on indicators for coach gender, artist gender, and the interaction term. Each specification includes coach and country fixed effects. Column 2 adds the order of the performance fixed effects. Column 3 includes the same controls as column 3, with the exception of adding genre-fixed effects. Column 4 adds controls for the number of males relative to females
in the team, the failure rate of the coach
for female artists, and the failure rate of the coach
for male artists. Standard errors are clustered by coach.
 * \(p<0.1\), ** \(p<0.05\), *** \(p<0.01\)}
\end{minipage}
\end{table}

The additional columns of Table \ref{tab:regressions}, address potential confounders to this gender bias estimate. Column (2) adds the order of the performance fixed effect to the coach fixed effect to account for order trends; results in these columns slightly increase in magnitude compared with those in Column (1). I have also included genre covariates in Column (3), which means coaches of a particular gender disproportionately favor certain genre styles. Suppose contestants of a different gender from these coaches are disproportionately likely to choose genre style. In that case, failure to control for genre covariates could result in a spurious gender bias estimate. In Column (4), I have incorporated the show-related specification, which I specify as constraints to the coaches. It comes from the coaches having a limited slot in their team, which will be a confounding factor in getting selected. For example, if a coach has a majority of women on the team, it might push the coach to change the strategy of choosing the contestant. Similarly, the coach has been rejected by a particular gender disproportionally more, which might lead to reluctance in selecting the next same-gender contestant. By including these covariates as confounding factors, I can control for constraints imposed by this contested environment. Across these five columns, the female/male estimates range from 4.4 to 5 percentage points. These four estimates are statistically significant at \(p<0.05\).

To sum up, the primary result of this study is that there is meaningful opposite-gender bias by coaches in a context in which the performance of contestants is exogenously assigned to the coaches. This finding contrasts significantly with the previous work of \citet{Carlsson2019}. They found that women (female recruiters or firms with a high share of female employees) favor women in the recruitment process. However, they did not see much evidence that men (male recruiters or firms with a high share of male employees) favor men. Results could differ due to differences in the sample studied. As the author analyzes low- and medium-skilled occupations in Sweden, there could be differences between the countries I use and Sweden. To make the comparison, I split my sample into separate countries and compared the results with Table \ref{tab:regressions}. The reported coefficients for the UK, France, and Australia in Table \ref{tab:regressions1} provide a similar magnitude to my initial analysis but with less precision. The standard errors are high due to the decreased sample size. Regardless, Germany has a null coefficient. Thus, there could be differences in the results due to the context of the country. Again, I account for this discrepancy in my initial analysis by controlling for country-fixed effects. The remainder of this manuscript examines heterogeneity in these effects to better characterize the nature of this gender bias and shed light on the likely mechanisms.

\subsection{Effect of Own and Opposite Gender Coaches by Constraints \label{sec:heterogeniety1}}

Given the findings discussed above, it is natural to ask whether the differences by gender are due to only a handful of coaches or if they are more systemic. To address this question, the remainder of the empirical results examines heterogeneity in the estimated gender bias by confounding factors - team gender composition, order of performance, and failure rates of the coaches. We also discuss the extent to which difference-in-difference estimates are sensitive to these confounding factors. To do so, I use the causal forest approach described in Section 4 to estimate the heterogeneous effects of gender bias.

Because the difference in differences estimator can only identify relative gender bias across coaches, it does not show gender bias separately for male and female coaches. There could be a case when the likelihood of male coaches preferring female artists is more than female coaches to female artists. If we decompose this into separate female and male coaches, male coaches prefer female artists more than male artists, and female coaches prefer female artists more than male artists. In this case, difference-in-differences estimation shows an opposite gender bias, but opposite gender bias occurs for male coaches and own-gender bias for female coaches. To observe the source of the bias, I estimate gender bias separately for men and women for a more straightforward interpretation. 

\begin{figure}[H]
    \caption{Estimate of gender bias by gender composition in the team}
    \label{fig:gender_composition}
    \centering

	\subfloat[Conditional on Male Coach ]{
	    \includegraphics[width=3.0in]{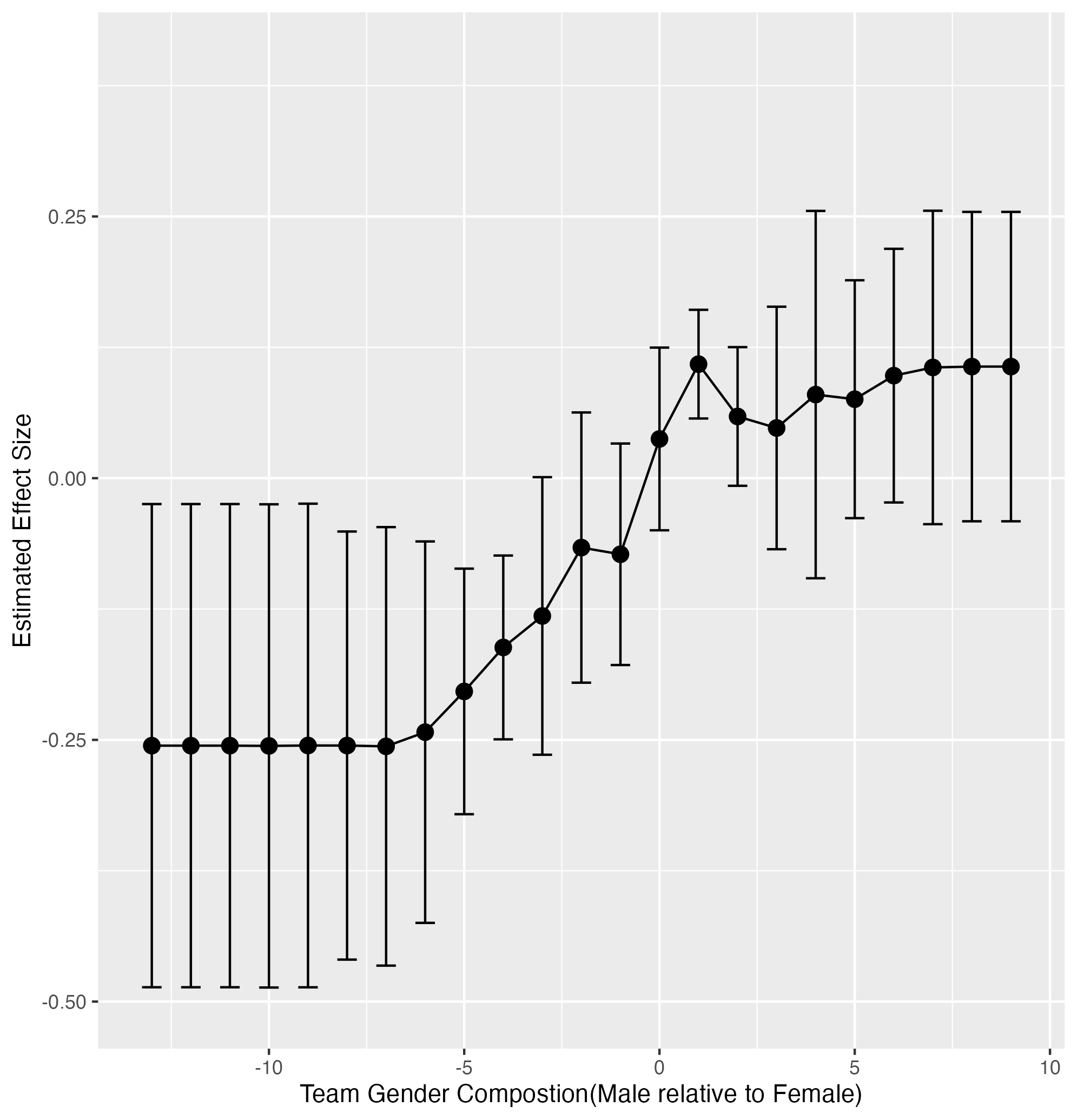}
	 }
	\subfloat[Conditional on Female Coach]{
	    \includegraphics[width=3.0in]{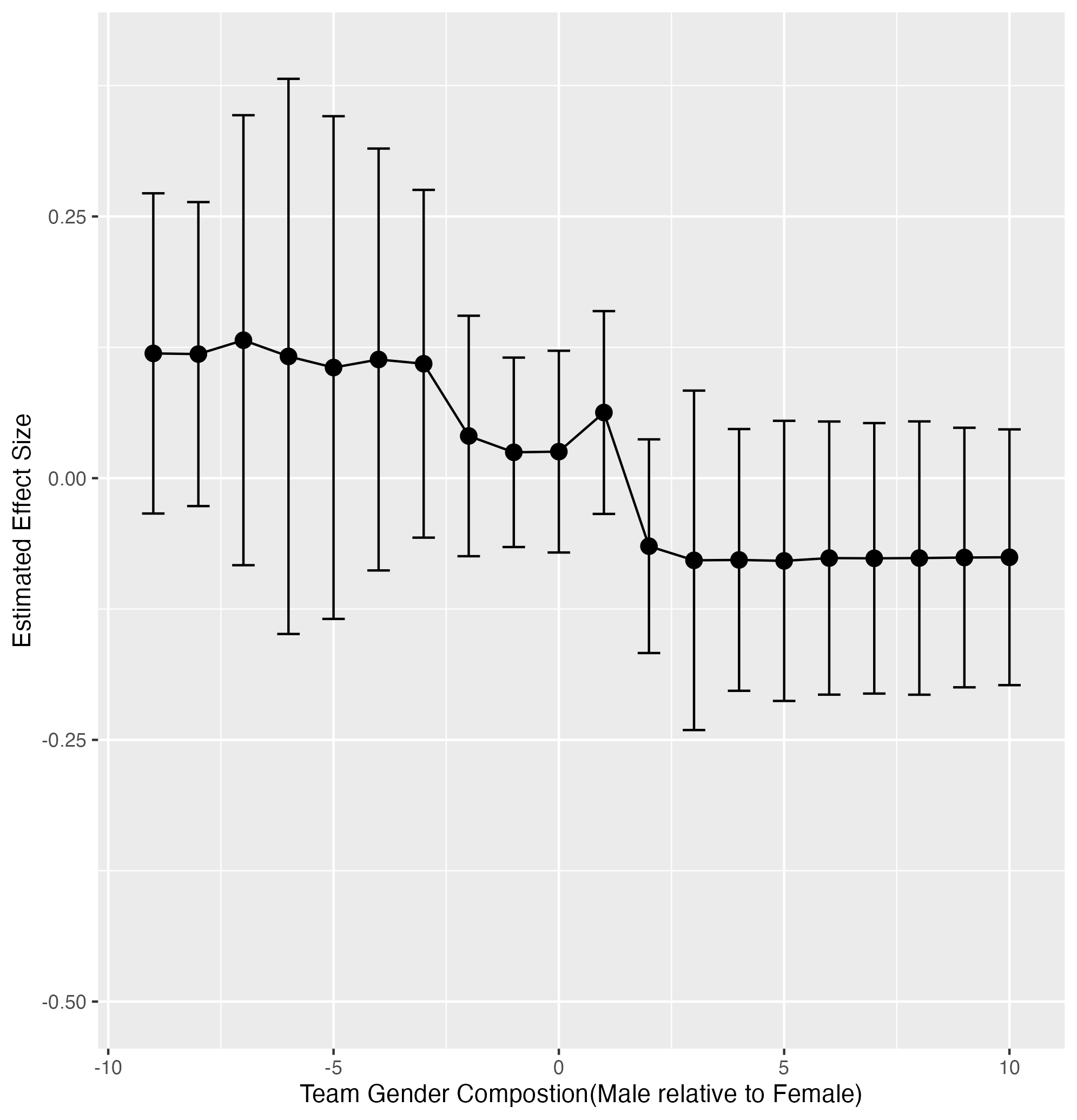}
	 }
    \vspace{0.2cm}
    \begin{minipage}{0.95\textwidth} 
	{\footnotesize Notes: This figure summarizes the results from the causal forest by approach in Athey and Wager (2019), where each estimate shown represents the effect of own-gender and opposite-gender interactions on selecting the artists in the case controlling for the gender of the coach and coach fixed effects. From left to right, the outcome of interest is a set of indicators for the differences in the number of male artists to the number of female artists in the team during the audition. This result effectively yields the estimates of own and opposite gender bias in selections that are specific to each team's gender composition. Standard errors clustered by coach.
	}
	\end{minipage}
\end{figure}

To examine heterogeneity at the coach level, I fixed coach characteristics such as coach gender. Firstly, I evaluate heterogeneity in gender bias by the gender composition of the team. To do so, I take the difference between the number of men and women in the team when the coaches hear the performance. The results of this exercise are shown in Figure \ref{fig:gender_composition}, where the estimated effect translates into the percentage point of selecting a male artist. We can see specific local average treatment effect coefficients for gender bias when we vary team gender composition. If the majority gender group in the team is women, male coaches tend to select more female artists. 

In contrast, when a majority of the team is men, male coaches tend to choose male artists, unlike female coaches, who have the opposite tendency to select opposite-gender artists when a particular gender is in the minority. The exciting aspect of this analysis is that the estimated effect on the likelihood of choosing male artists is essentially centered to zero when the number of men and women in the team is equal. 

I have also done the same heterogeneity analysis of gender team composition by standard ordinary least squares by grouping my dataset into males relative to the female difference in each coach's team during the audition. According to Table \ref{tab:regressions2}, the estimate shows a statistically significant result when the teams are female-dominant, and the magnitude is around 8.5 percentage points of the opposite gender bias. The weight for this estimate comes from the male coach being more likely to select female artists. The analysis provides a consistent result with a causal forest estimation but has little statistical power because of the data subsets by certain cutoffs.

This exercise yields additional compelling evidence that gender bias, I estimate, is a systematic effect not attributable to just a handful of coaches or types of situations. Similarly, I ask whether estimates are driven by failure rates of coaches for male and female artists and the order of the artists' performance. It is thus unsurprising to see in Figures \ref{fig:failure_male}, \ref{fig:failure_female}, and \ref{fig:order} that other variable-specific estimates of gender bias are uncorrelated with the failure rate of coaches for male and female artists and the order of the performance.

\section{Discussion and Conclusion\label{sec:conclusion}}

In this paper, we examine whether the gender of the coach and artist matters when it comes to selecting the performance in the blind audition stage in the Voice TV show. Results provide strong evidence that gender matters and I show that coaches choose other gender artists 12 percent more when they hear similar performances in the audition. The results are robust with many variations, such as coach and artist-specific covariates. Under the interpretation of opposite-gender bias, the results would be due to male and/or female coaches being either too lenient to opposite-gender artists, being too tough on own-gender artists, or both.  

Relatedly, a potential mechanism through which opposite-gender interaction cause statistically significant gender bias could be employers hiring for female-dominated occupations showing a preference for male applicants \citep{Doris2004, Protsch2021}. However, \citet{Doris2004} and \citet{Protsch2021}'s research also found that employers preferred male applicants in male-dominated occupations. Still, because of data limitations, they do not observe the gender of the recruiter. Others who do not observe what I observe in my work found that employers prefer male applicants in male-dominated occupations and female applicants in female-dominated occupations \citep{Campero2018, Benard2010, Glick1988, Ridgeway2004}. These works provide robust evidence of gender bias in hiring. Still, we cannot distinguish whether it is an opposite-gender or own-gender bias hiring since the gender of the decision-makers is unobservable or decisions are made collectively. In my setting, I observe the gender of the coach and the coach's individual choice in selecting the artists. Therefore I incorporate this gender composition feature into my setting to address whether the coach changes their behavior based on the gender composition in the team. For example, If one coach's team is dominated by one gender, would the coach shift behavior during the audition and select an artist from another gender, a minority gender?

I conduct a heterogeneity analysis assessing the effect of different gender compositions in the team on the likelihood of being selected. To do so, I took advantage of the fact that the sample I used includes both selection of male and female coaches in the blind audition stages and if the artist ended up being in the team of the selected coach. Additionally, the other advantage of the sample is substantial variations in gender composition in each team, including both male-dominated and female-dominated settings. Typically, we observe only one type of setting, where the firm is female-dominated or male-dominated. Leveraging this variation in the gender composition, I estimate the heterogeneous effect of gender bias for male and female coaches. I found that male artists have an advantage over female artists in a male-dominated team when the coach is male. Likewise, women have an advantage over men in a female-dominated team. When the coach is female, we can see opposite scenarios. Female coaches tend to choose women over men in male-dominated teams and men over women in the female-dominated team. This difference is partially statistically significant, but the magnitude remains consistent. 

The literature provides robust support for a hypothesis on group formation dynamics, which is different for male and female leaders. One study of bias in other contexts suggests that men might not respond favorably to the presence of gender diversity, particularly in domains that men have historically dominated. Also, gender-related issues were most likely to be raised in groups where women are a minority or where the female is a leader \citep{CROCKER1984}. In addition, there is substantial evidence that women in team formation care about equality, men care about efficiency, and men consider male or female dominance in the team as more efficient. More directly, \citet{Kuhn2014} found that women respond differently to alternative rules for team formation in a manner consistent with advantageous inequity aversion. In contrast, men show greater responsiveness to efficiency gains associated with team production, meaning that when forming a team, men make their selections based on who will be the most efficient rather than on the grounds of gender equality. Overall, our findings indicate that gender composition and the gender of the decision-maker are important determinants in the hiring process. While it is difficult to know if these findings extend to other industries, these results corroborate the debate over gender discrimination in the hiring process. 

In conclusion, the results add evidence to a growing literature testing own or opposite gender bias in decision-making. This finding provides insights into the extent to which situational factors work to mitigate or exacerbate gender disparities in the hiring process. I show strong evidence of the heterogeneity in gender bias from gender composition in the team. This evidence suggests that the selection process is primarily rooted in the female and male recruiters' behavioral preferences of inequality or efficiency. In addition, the findings offer a new perspective to enrich past research on gender discrimination, shedding light on the instances of gender bias variation by the gender of the decision maker and team gender composition.

\clearpage
\begin{singlespace}
\bibliographystyle{aer}
\bibliography{our-cites.bib}
\end{singlespace}

\newpage
\appendix
\setcounter{table}{0}
\renewcommand{\tablename}{Appendix Table}
\renewcommand{\figurename}{Appendix Figure}
\renewcommand{\thetable}{A\arabic{table}}
\setcounter{figure}{0}
\renewcommand{\thefigure}{A\arabic{figure}}

\section{Appendix Tables and Figures}
\begin{table}[H]\centering
\def\sym#1{\ifmmode^{#1}\else\(^{#1}\)\fi}
\caption{Estimates of own-gender bias in artist selection by country}
\begin{tabular}{l*{5}{c}}
\toprule
&\multicolumn{1}{c}{(1)}&\multicolumn{1}{c}{(2)}&\multicolumn{1}{c}{(3)}&\multicolumn{1}{c}{(4)}\\
&\multicolumn{1}{c}{\specialcell{Germany}}&\multicolumn{1}{c}{\specialcell{Australia}}&\multicolumn{1}{c}{\specialcell{UK}}&\multicolumn{1}{c}{\specialcell{France}}\\
\midrule

\multicolumn{4}{l}{\textbf{Outcome: Choice}} \\
Male Coach*Male Artist &     0.0168	& -0.0473 &	-0.0687 &	-0.0531

  \\
  & (0.0455) &	(0.0366) &	(0.0425) &	(0.0386)
    \\
  \\
Male Coach & -0.0614\sym{**} &	-0.0990\sym{***} &	0.0617 &	-0.124\sym{***}\\
 & (-0.0247) &	(-0.0326) &	(-0.0408) &	(-0.0189)
 \\
 \\
 Male Artist & -0.00171 & 0.0559\sym{*} &	-1.60E-05 &	0.0397\\

& (0.0413) &	(0.0281) &	(0.0388) &	(0.0255)\\
\\
Mean       &     0.411         &     0.422          &   0.301  & 0.446   \\
$N$          & 2,688 &	2,964 &	2,860 &	3,396
  \\
  \\
  Coach FE       &     Y         &     Y          &   Y  &  Y    \\
 
Performance Order FE          & Y & Y & Y & Y
  \\
  Genre FE          & Y & Y & Y & Y\\
  Constraints  & Y & Y & Y & Y
  \\
\bottomrule
\end{tabular}
\label{tab:regressions1}
\vspace*{0.09cm}
\begin{minipage}{0.95\textwidth} 
{\footnotesize 
Notes: Table A1 presents a difference in differences estimate of gender bias using regression of Equation 1. This table reports the coefficient on the interaction of Male Artist and Male Coach from the regression of Choice on indicators for coach gender, artist gender, and the interaction term. The reported coefficients are for "own-gender coach" using data subsets by country indicated in columns. Each specification includes coach and country fixed effects, the order of the performance fixed effects, genre-fixed effects, the number of males relative to females in the team, the failure rate of the coach for female artists, and the failure rate of the coach for male artists. Standard errors are clustered by coach.
 \sym{*} \(p<0.1\), \sym{**} \(p<0.05\), \sym{***} \(p<0.01\)}
\end{minipage}
\end{table}

\begin{table}[H]\centering
\def\sym#1{\ifmmode^{#1}\else\(^{#1}\)\fi}
\caption{Estimates of own gender-race bias in artist selection by team gender composition}
\begin{tabular}{l*{5}{c}}

\toprule
&\multicolumn{1}{c}{(1)}&\multicolumn{1}{c}{(2)}&\multicolumn{1}{c}{(3)}&\multicolumn{1}{c}{(4)}\\

& &\multicolumn{2}{c}{Male relative to Female in the Team}& &\\
  & \(X< -2\) &	-\(3<X\)\(<0\)   &	-\(1<X\)\(<3\) &	\(X>3\)\\
\midrule
\multicolumn{4}{l}{\textbf{Outcome: Choice}} \\
Male Coach*Male Artist &     -0.0863\sym{**}	& -0.0852\sym{**} &	-0.0403 &	0.0519

  \\
  
  & (0.0417) &	(0.0366) &	(0.0293) &	(0.0661)
    \\
  \\
  
Male Coach  & -0.0109 &	-0.226\sym{***} &	-0.228\sym{***} &	0.633\sym{***}\\

& (0.0377)	& (0.0289) &	(0.0343) &	(0.136)\\
 
Male Artist & -0.108\sym{***} &	-0.0253	& 0.107\sym{***} &	0.0336\\
 & (0.0352)	& (0.0272) &	(0.0222) &	(0.0571)
 \\
 \\

\\
Mean       &     0.389         &     0.401         &   0.408  & 0.345    \\
$N$     &     2,174 &	3,304 &	4,469 &	1,188
  \\
  \\
  Coach FE       &     Y         &     Y          &   Y  &  Y    \\
 Country FE       &     Y         &     Y          &   Y  &  Y   \\
Performance Order FE          & Y & Y & Y & Y
  \\
  Genre FE          & Y & Y & Y & Y\\
  Constraints  & Y & Y & Y & Y
  \\
\bottomrule
\end{tabular}
\label{tab:regressions2}
\vspace*{0.09cm}
\begin{minipage}{0.95\textwidth} 
{\footnotesize 
Notes: Table A2 presents a difference in differences estimate of gender bias using regression of Equation 1. This table reports the coefficient on the interaction of Male Artist and Male Coach from the regression of Choice on indicators for coach gender, artist gender, and the interaction term. The reported coefficients are for "own-gender coach" using data subsets by gender composition in the team indicated in columns. \(X< -2\) means that difference in the number of women and men in the team is 3 or higher. Similarly, \(X>3\) means that difference in the number of men and women is 4 or higher. Each specification includes coach and country fixed effects, the order of the performance fixed effects, genre-fixed effects, the failure rate of the coach for female artists, and the failure rate of the coach for male artists. Standard errors are clustered by coach.
 \sym{*} \(p<0.1\), \sym{**} \(p<0.05\), \sym{***} \(p<0.01\)}
\end{minipage}
\end{table}

\begin{figure}[H]
    \caption{Estimate of gender bias by failure rate for male artists }
    \label{fig:failure_male}
    \centering

	\subfloat[Conditional on Male Coach ]{
	    \includegraphics[width=3.0in]{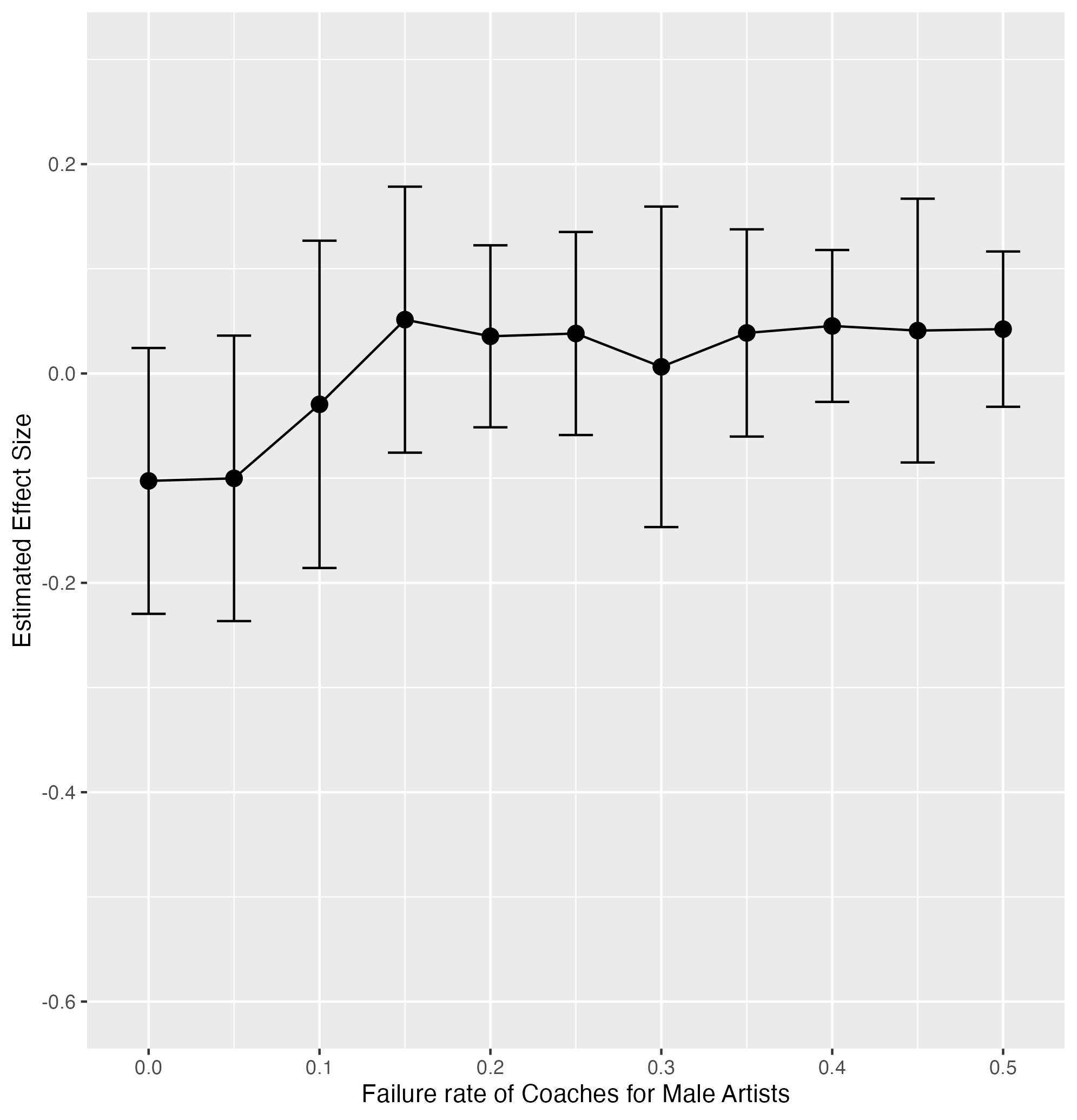}
	 }
	\subfloat[Conditional on Female Coach]{
	    \includegraphics[width=3.0in]{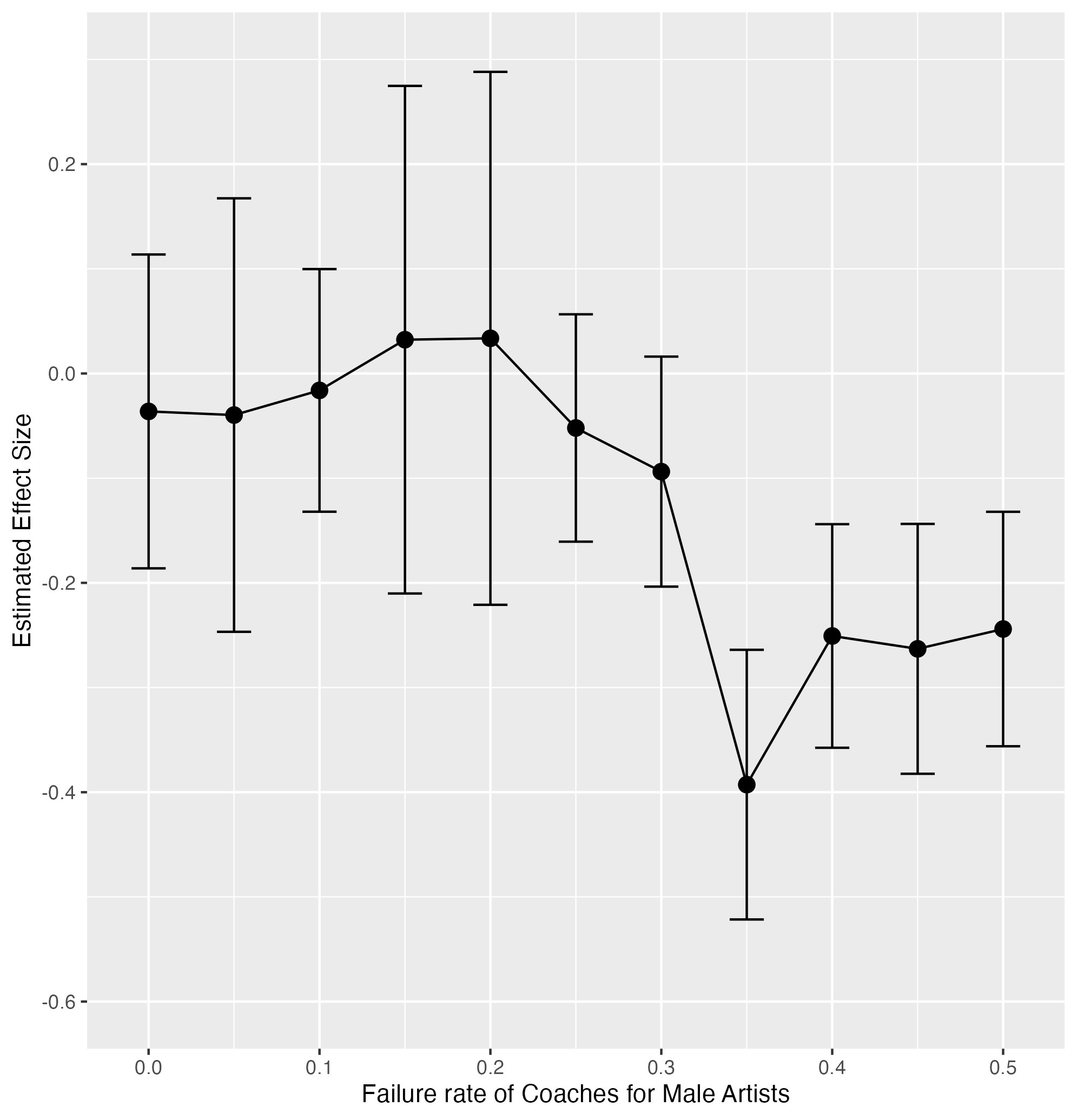}
	 }
    \vspace{0.2cm}
    \begin{minipage}{0.95\textwidth} 
	{\footnotesize This figure summarizes the results from the causal forest by approach in Athey and Wager (2019), where each estimate shown represents the effect of own-gender and opposite-gender interactions on selecting the artists in the case controlling for the gender of the coach and coach fixed effects. From left to right, the outcome of interest is a set of indicators for the failure rate of coaches for male artists. This is the rate at which a coach wanted an artist of a particular gender on their team, and the artist did not choose the coach. This result effectively yields the estimates of own and opposite gender bias in selections that are specific to variation by the failure rate of coaches for male artists. Standard errors clustered by coach.
	}
	\end{minipage}
\end{figure}

\begin{figure}[H]
    \caption{Estimate of gender bias by failure rate for female artists}
    \label{fig:failure_female}
    \centering

	\subfloat[Conditional on Male Coach ]{
	    \includegraphics[width=3.0in]{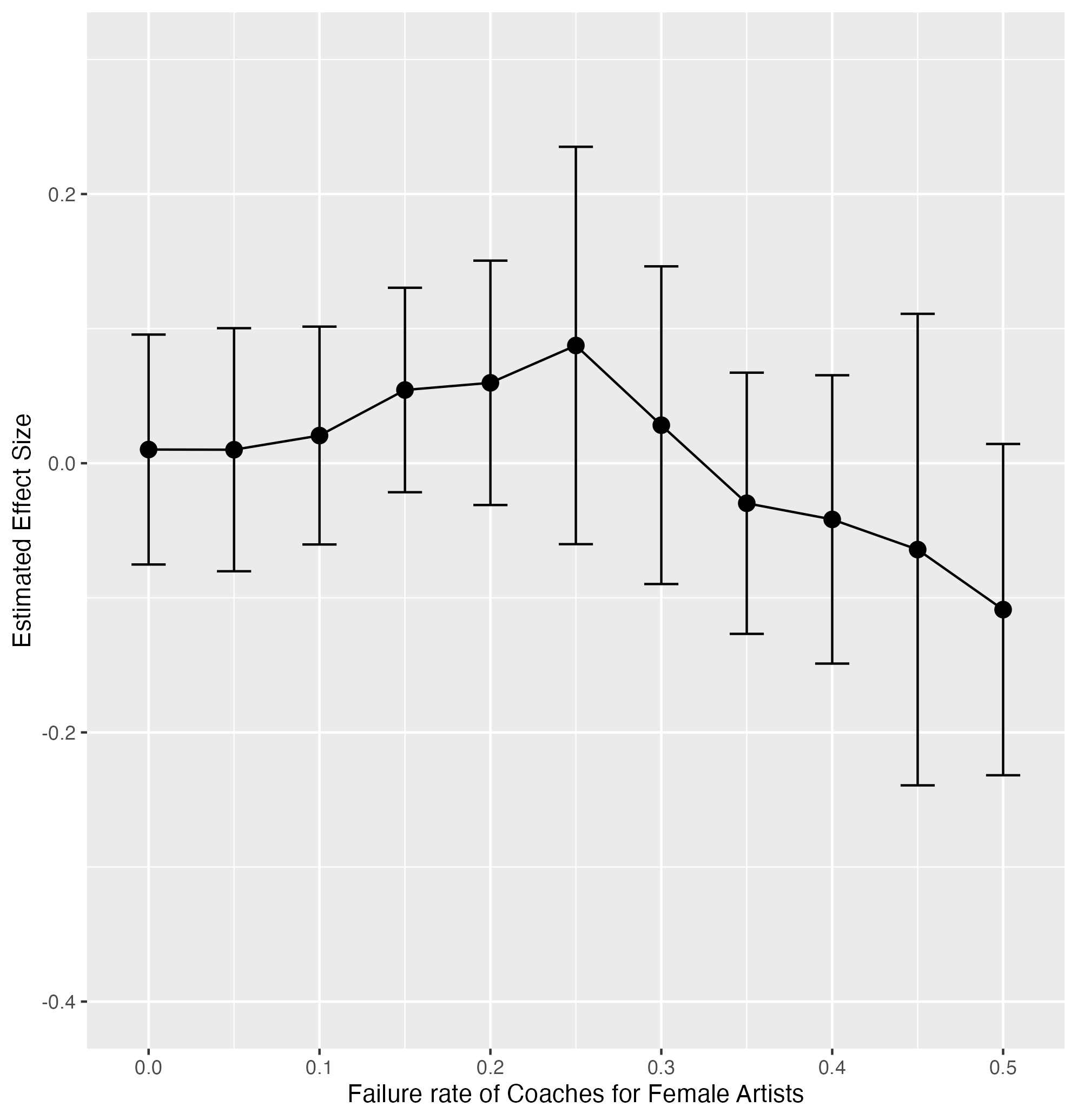}
	 }
	\subfloat[Conditional on Female Coach]{
	    \includegraphics[width=3.0in]{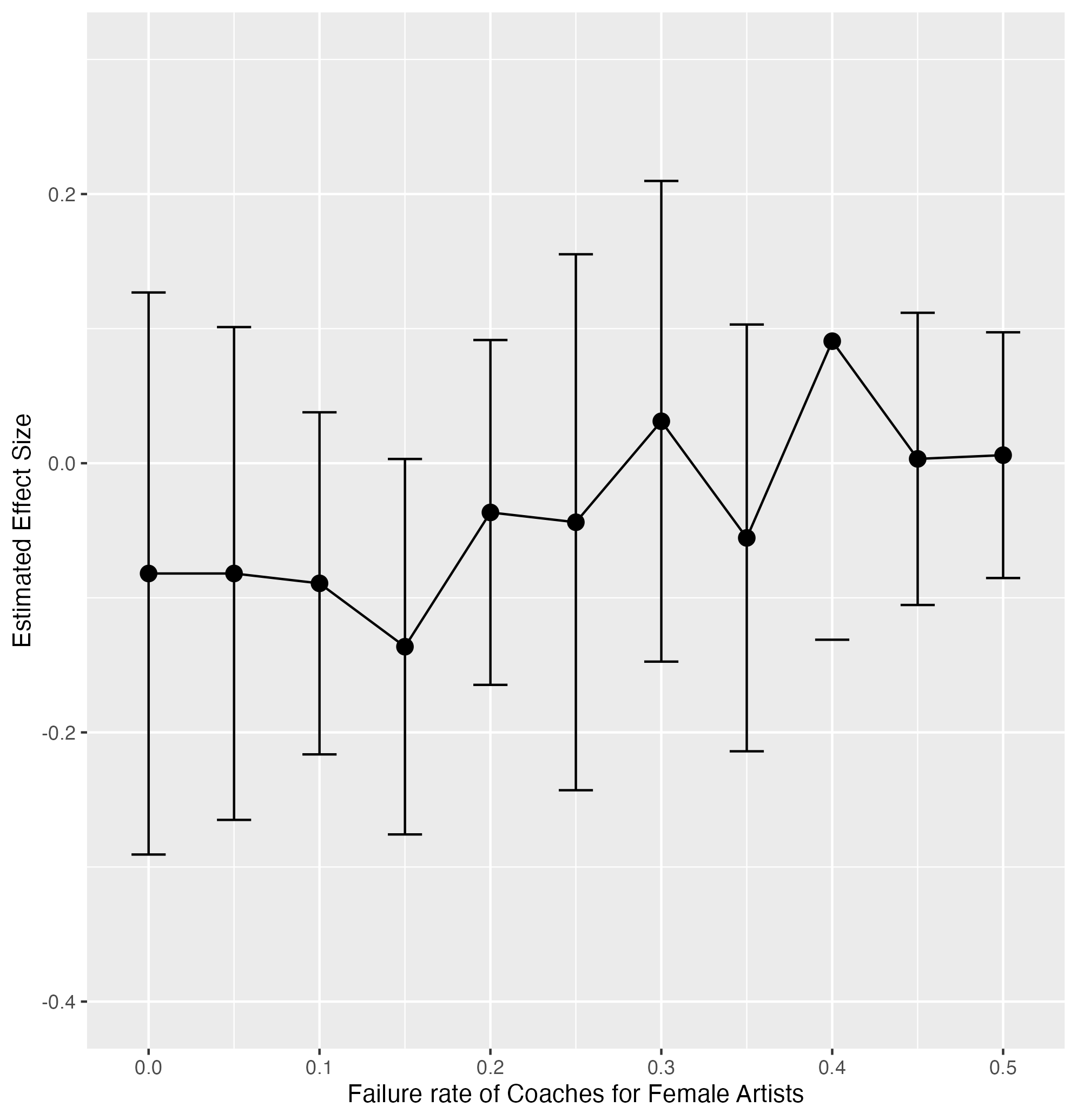}
	 }
    \vspace{0.2cm}
    \begin{minipage}{0.95\textwidth} 
	{\footnotesize This figure summarizes the results from the causal forest by approach in Athey and Wager (2019), where each estimate shown represents the effect of own-gender and opposite-gender interactions on selecting the artists in the case controlling for the gender of the coach and coach fixed effects. From left to right, the outcome of interest is a set of indicators for the failure rate of coaches for female artists. This is the rate at which a coach wanted an artist of a particular gender on their team, and the artist did not choose the coach. This result effectively yields the estimates of own and opposite gender bias in selections that are specific to variation by the failure rate of coaches for female artists. Standard errors clustered by coach.
	}
	\end{minipage}
\end{figure}

\begin{figure}[H]
    \caption{Estimate of gender bias by the performance order}
    \label{fig:order}
    \centering

	\subfloat[Conditional on Male Coach ]{
	    \includegraphics[width=3.0in]{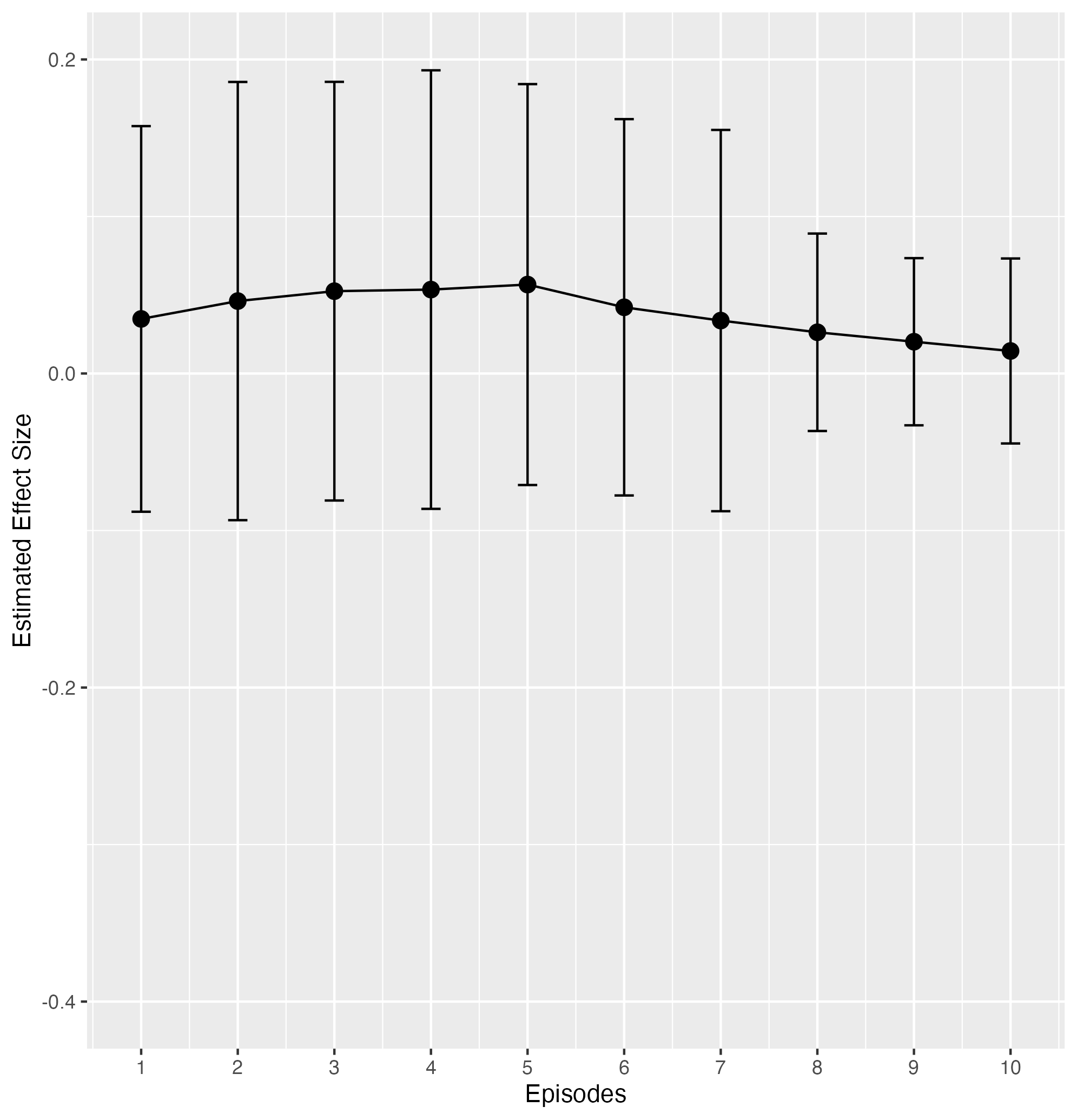}
	 }
	\subfloat[Conditional on Female Coach]{
	    \includegraphics[width=3.0in]{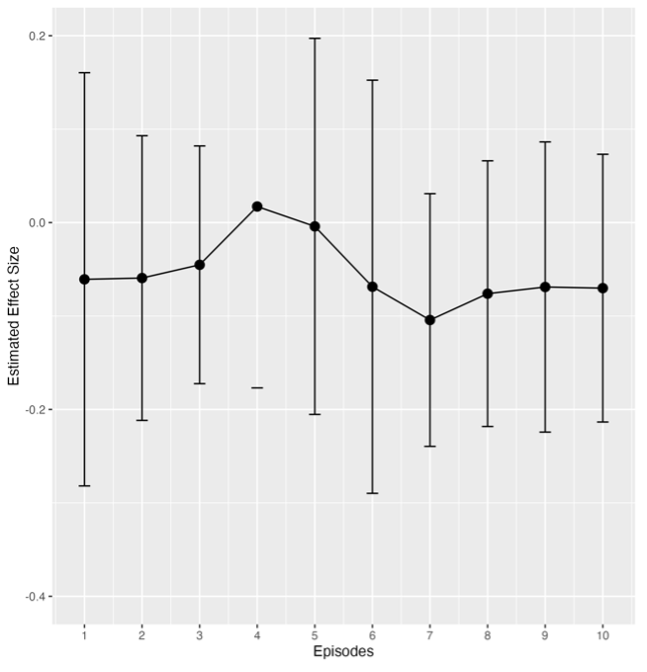}
	 }
    \vspace{0.2cm}
    \begin{minipage}{0.95\textwidth} 
	{\footnotesize This figure summarizes the results from the causal forest by approach in Athey and Wager (2019), where each estimate shown represents the effect of own-gender and opposite-gender interactions on selecting the artists in the case controlling for the gender of the coach and coach fixed effects. From left to right, the outcome of interest is a set of indicators for each audition episode. Each 10th order of ten performances aggregates into the episodes in the same order. This result effectively yields the estimates of own and opposite gender bias in selections that are specific to variation by the grouped order of performance. Standard errors clustered by coach.
	}
	\end{minipage}
\end{figure}

\end{document}